\documentclass[12pt,a4paper,oneside,notitlepage]{article}
\newif\ifpdf
\ifx\pdfoutput\undefined
   \pdffalse
\else
  \pdfoutput=1
  \pdftrue
\fi

\ifpdf
  \usepackage[pdftex]{graphicx}
  \usepackage[pdftex]{hyperref} 
\else
  \usepackage{graphicx}
  \usepackage{hyperref} 
\fi

\usepackage{c-series}

\usepackage{url}
\usepackage{amsfonts}

\usepackage{theorem}

\newtheorem{theorem}{Theorem}
\newtheorem{lemma}{Lemma}
\newtheorem{corollary}{Corollary}
\newenvironment{proof}{\emph{Proof. }}{\hfill $\Box$\vspace{2ex}

}

\newcommand{\true}{\textsc{true}}
\newcommand{\false}{\textsc{false}}

\begin{document}
%\renewcommand{\baselinestretch}{1.24}
%\large
%\normalsize

%
% Identification info
%

\title{The Complexity of Maximum Matroid-Greedoid Intersection and
Weighted Greedoid Maximization}
\author{Taneli Mielik\"ainen \and Esko Ukkonen}
\authorcontact{Taneli Mielik\"ainen \\
\texttt{tmielika@cs.Helsinki.FI}
\and Esko Ukkonen \\
\texttt{ukkonen@cs.Helsinki.FI}}
\titlelinebreaks{The Complexity of Maximum \\
Matroid-Greedoid Intersection and \\
Weighted Greedoid Maximization}

\reportyear{2004}
\reportno{2}
\reportmonth{February}
\CRClasses{F.2.2,G.2.1}
\printhouse{}
\reportpages{\pageref{contentsendpage} + \pageref{lastpage}}

\CRClassesLong{\newlength{\restline}
\setlength{\restline}{\textwidth}
\addtolength{\restline}{-13mm}
\parbox[t]{13mm}{F.2.2}
\vspace{1ex}
\parbox[t]{\restline}{Analysis of Algorithms and Problem Complexity: Nonnumerical Algorithms and Problems: Computations on discrete structures}
}               

\GeneralTerms{Algorithms, Theory}
\AdditionalKeyWords{Combinatorial Optimization, $NP$-Hardness, Inapproximability, Fixed-parameter Intractability}

%%%%%%%%%%%%%%%%%%%%%%%%%%%%%%%%%%%%%%%%%%%%%%%%%%%%%%%%%%%%%%%%%%%%%%%%%%%%%

%
% Cover, title and abstract pages for C-series publication
%

\pagestyle{empty}

% cover page with HY logo
\makecover

% address, email, etc.
%\makecontactsheet

% inner title page
%\maketitlepage

% abstract
\cleardoublepage
%\clearpage
\abstractpagestart
%\begin{abstract}
The maximum intersection problem for a matroid and a greedoid, given
by polynomial-time oracles, is shown $NP$-hard by expressing the
satisfiability of boolean formulas in $3$-conjunctive normal form as
such an intersection.  The corresponding approximation problems are
shown $NP$-hard for certain approximation performance bounds.
Moreover, some natural parameterized variants of the problem are shown
$W[P]$-hard.  The results are in contrast with the maximum
matroid-matroid intersection which is solvable in polynomial time by
an old result of Edmonds.  We also prove that it is $NP$-hard to
approximate the weighted greedoid maximization within $2^{n^{O(1)}}$
where $n$ is the size of the domain of the greedoid.

A preliminary version ``The Complexity of Maximum Matroid-Greedoid
Intersection'' appeared in Proc.\ FCT 2001, LNCS 2138, pp.\ 535--539,
Springer-Verlag 2001.
%\end{abstract}
\abstractpageend

% table of contents
\clearpage
\pagestyle{headings}
%\thispagestyle{empty}
%\tableofcontents
\label{contentsendpage}

%%%%%%%%%%%%%%%%%%%%%%%%%%%%%%%%%%%%%%%%%%%%%%%%%%%%%%%%%%%%%%%%%%%%%%%%%%%%%

%
% Document body
%

% restart page numbering using arabic numbers
\newpage
\pagenumbering{arabic}
\setcounter{page}{1}
\thispagestyle{plain}

% lower-case headers
\renewcommand{\sectionmark}[1]{%
\markright{\sectionname
\ \thechapter.\ #1}{}%
}

\pagestyle{headings}

% input chapters like below

%%%%%%%%%%%%%%%%%%%%%%%%%%%%%%%%%%%%%%%%%%%%%%%%%%%%%%%%%%%%%%%%%%%%%%%%%%%%
\section{Introduction}
A set system $(S,F)$ where $S$ is a finite set (the \emph{domain} of
the system) and $F$ is a collection of subsets of $S$ is a
\emph{matroid} if
\begin{enumerate}
\item[]
\begin{enumerate}
\item[(M1)] $\emptyset \in F$;
\item[(M2)] If $Y \subseteq X \in F$ then $Y \in F$;
\item[(M3)] If $X,Y \in F$ and $|X|>|Y|$ then there is an $x \in X
\setminus Y$ such that $Y \cup \left\{x \right\} \in F$.
\end{enumerate}
\end{enumerate}
A \emph{greedoid} is a set system $(S,F)$ that satisfies (M1) and
(M3).

In applications a matroid or a greedoid is given by an oracle, i.e.,
by a deterministic algorithm that answers the question whether $X$
belongs to $F$ for any $X \subseteq S$.

Many combinatorial problems can be formulated using matroids or
greedoids (see e.g. \cite{kv-co,ps-co}).  The seminal example is the
maximum matching problem in bipartite graphs.  Each instance of the
problem can be represented as the intersection of two matroids.  For a
bipartite graph $B=\left(V \cup V',E\right)$ where $V \cap
V'=\emptyset$ and $E \subseteq V \times V'$, the first matroid
consists of all subsets of the edge set $E$ such that a subset
contains at most one edge starting from the same node in $V$.  The
second matroid consists of similar subsets but now a subset can
contain at most one edge ending at the same node in $V'$.  Then the
maximum matching corresponds to the largest set in the intersection of
the two matroids.

We want to consider in the matroid-greedoid framework the
computational complexity of general combinatorial problems that have
infinitely many instances. Therefore we introduce families of matroids
and greedoids that have uniform polynomial-time representations as
follows. Let $\mathcal{F}=\left\{ (S_h,F_h)_{h \in H} \right\}$ be a
possibly infinite set of matroids or greedoids.  Then $\mathcal{F}$ is
said to be given by a \emph{uniform polynomial-time oracle} if there
is an algorithm $\mathcal{O}$, that when given $h$ and some $X
\subseteq S_h$ answers whether or not $X \in F_h$ in time polynomial
in $|S_h|$.

Let $\mathcal{F}=\left\{ (S_h,F_h)_{h \in H} \right\}$ and
$\mathcal{G}=\left\{ (S_h,G_h)_{h \in H} \right\}$ be two such
families given by uniform polynomial-time oracles.  Note that the
index set $H$ is the same for both, and for a given $h$, both have the
same domain $S_h$.

The \emph{maximum intersection problem} for $\mathcal{F}$ and
$\mathcal{G}$ is to find, given an index $h \in H$, a set $X \in F_h
\cap G_h$ such that $|X|$ is maximum.  A solution algorithm of the
maximum intersection problem is polynomial-time if its running time is
polynomial in $\left|S_h\right|$.

Edmonds \cite{e-mpm} gave the first polynomial-time solution for the
intersection problem in the case that both $\mathcal{F}$ and
$\mathcal{G}$ are families of matroids.  In this paper we consider the
obvious next step, namely the intersection of families of matroids and
greedoids.

The following constrained version of the bipartite matching gives an
example of a problem that can be represented as an intersection of a
greedoid and a matroid.  The problem is called the \emph{maximum
tree-constrained matching problem}.  An instance of it consists of a
bipartite graph $B=\left(V \cup V',E\right)$ and a rooted tree
$T=\left(V,D,r\right)$ where $r \in V$ is the root.  The tree $T$ will
constrain the use of $V$ in the matching: the problem is to find a
maximum-size matching in $B$ such that the \emph{matched} nodes of $V$
include $r$ and induce a connected subgraph (actually a tree rooted at
$r$) of $T$.

To represent this problem as an intersection of a greedoid and a
matroid we modify the construction given above for the unconstrained
bipartite matching.  The collection of subsets of edges ending at
different nodes in $V'$ remains as in the unconstrained case.  Hence
it is a matroid.  The collection of subsets of edges starting from
different nodes in $V$ must now satisfy the additional tree-constraint
given by $T$.  That is, this collection only contains edge sets such
that each subset $W$ of $V$ that is adjacent to such an edge set
contains the root node $r$ and forms a connected subgraph of $T$.  It
follows straightforwardly from the properties of connected subgraphs
of a tree that this collection is a greedoid (but not necessarily a
matroid).  It is immediate that the largest element in the
intersection of the greedoid and the matroid is a solution of our
maximum tree-constrained matching problem.

A closer look also reveals that the difficulty of the problem is
determined by the topology of the tree $T$ or, more precisely, by the
number of connected subgraphs of $T$ that contain $r$.  If the number
of such subgraphs is polynomial in $\left|V\right|$ (this is the case
for example if $T$ is a path), then the maximum tree-constrained
matching can be found in polynomial time: we just apply Edmond's
algorithm repeatedly on all bipartite graphs that are obtained from
$B=\left(V \cup V',E\right)$ by replacing $V$ with the nodes $W
\subseteq V$ in each connected subgraph of $T$.  The largest matching
found that matches the corresponding $W$ entirely is a solution of our
problem.  The number of connected subgraphs can be super-polynomial
(for example if $T$ is a balanced binary tree) suggesting that our
problem might not be polynomial-time solvable in general.
Consistently with this observation we will show that the maximum
matroid-greedoid intersection problem is $NP$-hard.

The paper has been organized as follows.  In Section~\ref{s:nph} we
show, by reduction from 3SAT, that the maximum intersection problem
for a matroid family and a greedoid family, given by uniform
polynomial-time oracles, is $NP$-hard.  In Section~\ref{s:approx},
this reduction is modified to show that the maximum matroid-greedoid
intersection problem is not approximable within a factor
$|S_h|^{1-\epsilon}$ for any fixed $\epsilon>0$, and its weighted
version, the maximum weight matroid-greedoid intersection problem, is
not approximable within $2^{|S_h|^k}$ for any fixed $k>0$, unless
$P=NP$.  Finally, in Section~\ref{s:fpt}, we consider our problem in
the parameterized complexity framework. We show that it is $W[P]$-hard
to decide whether or not a matroid-greedoid intersection contains a
set of given size.

\section{$NP$-hardness \label{s:nph}}

The hardness proofs in this section and in Section~\ref{s:approx}
reduce some $NP$-hard problem $H$ to the intersection problem of a
matroid family $\mathcal{F}$ and a greedoid family $\mathcal{G}$.
This polynomial-time many-one reduction is non-standard as
$\mathcal{F}$ and $\mathcal{G}$ are given only by oracles.  An
instance $h \in H$ will be reduced to $(S_h,F_h,G_h)$.  Here $S_h$ is
the domain, $(S_h,F_h)$ a matroid and $(S_h,G_h)$ a greedoid such that
the families $\mathcal{F}= \left\{ (S_h,F_h)_{h \in H} \right\}$ and
$\mathcal{G}= \left\{ (S_h,G_h)_{h \in H} \right\}$ are given by
uniform polynomial-time oracles.

The reduction step $h \mapsto (S_h,F_h,G_h)$ is implemented by
specializing the uniform oracle algorithms $\mathcal{O}$ and
$\mathcal{O}'$ for the matroid family $\mathcal{F}$ and the greedoid
family $\mathcal{G}$ to $h$, giving specialized algorithms
$\mathcal{O}\left(h\right)$ and $\mathcal{O}'\left(h\right)$.
Specializing simply means that an input parameter of the algorithms is
fixed to the given value $h$.  This reduction can obviously be
accomplished in polynomial time.  The specialized oracles
$\mathcal{O}\left(h\right)$ and $\mathcal{O}'\left(h\right)$ recognize
members of $F_h$ and $G_h$ in time polynomial in $|S_h|$.  To make
this a hardness proof we must also require that $|S_h|$ is polynomial
in $|h|$. Then the running times of the oracles
$\mathcal{O}\left(h\right)$ and $\mathcal{O}'\left(h\right)$ actually
become polynomial in $|h|$.

Recall that the $NP$-complete problem \emph{$3$-satisfiability} (3SAT)
is, given a boolean formula $h$ in $3$-conjunctive normal form (3CNF),
to decide whether or not there is a truth assignment $Z$ for the
variables of $h$ such that $h(Z)=\true$.

We construct the instance $(S_h,F_h),(S_h,G_h)$ of matroid-greedoid
intersection that corresponds to $h$ as follows. Let $h$ contain $n$
different boolean variables.  Then $S_h$ contains symbols $t_1,f_1,
\ldots ,t_n,f_n$.  The symbols $t_1,f_1, \ldots ,t_n,f_n$ will be used
to encode truth assignments: $t_i$ encodes that the $i$th variable is
$\true$ and $f_i$ that it is $\false$.

The subset collection $F_h$ consists of all subsets of $S_h$ that
contain at most one of the symbols $t_i,f_i$ for $i=1,\ldots ,n$.  It
is immediate, that $(S_h,F_h)$ satisfies the matroid properties (M1),
(M2), and (M3).

The subset collection $G_h$ consists of two groups.  The first group A
consists of all subsets $X$ of $S_h$ such that $|X| \leq n$ and $X
\cap \left\{ t_n,f_n \right\} = \emptyset$.  The second group B
consists of the sets that represent a truth assignment that satisfies
$h$.  Such a set is of size $n$ and contains one element from each
$t_i,f_i$.

To verify that $(S_h,G_h)$ is a greedoid, first note that (M1) is
obviously true. To verify (M3), let $X,Y \in G_h$ such that $|X|>|Y|$.
\begin{enumerate}
\item
If $|X|<n$ then $X$ and $Y$ must belong to group A.  Hence for any
element $x \in X \setminus Y$, set $Y \cup \left\{ x \right\}$ belongs
to group A and hence to $G_h$.
\item
If $|X|=n$ and $|X \setminus Y|=1$ then $Y \cup (X \setminus Y)=X$,
i.e., property (M3) holds.
\item
In the remaining case $|X|=n$ and $|X \setminus Y|>1$.  As $X
\setminus Y$ contains at least two elements and no set of $G_h$
contains both $t_n$ and $f_n$, at least one element $x \in X \setminus
Y$ must be different from $t_n,f_n$.  Then $Y \cup \left\{ x \right\}$
belongs to group A.
\end{enumerate}

The matroid-greedoid intersection $F_h \cap G_h$ contains a set $X$
such that $|X|=n$ if and only if the group B in the definition of
$G_h$ is non-empty, that is, if and only if $h$ is satisfiable.  As
such a set $X$ is also the largest in $F_h \cap G_h$, we have shown:
\begin{lemma}
\label{l:bfmgi}
Boolean formula $h$ is satisfiable if and only if the maximum element
in $F_h \cap G_h$ for matroid $(S_h,F_h)$ and greedoid $(S_h,G_h)$ is
of size $n$ where $n$ is the number of variables of $h$.
\end{lemma}

The above construction yields a matroid family $\mathcal{F}=\left\{
(S_h,F_h)_{h \in \mathrm{3CNF}} \right\}$ and a greedoid family
$\mathcal{G}=\left\{ (S_h,G_h)_{h \in \mathrm{3CNF}} \right\}$.  Both
have a uniform polynomial-time oracle for checking membership in $F_h$
and $G_h$: The only nontrivial task of the oracle is to verify when a
truth assignment satisfies a given formula $h$, but this is doable in
time polynomial in $|h|$ using well-known techniques.  As $|h|=O(n^3)$
for a 3CNF formula $h$ and $|S_h|=2n$, the running time of the oracle
is polynomial in $|S_h|$, too.

It follows from Lemma~\ref{l:bfmgi} and the discussion above that our
construction is a polynomial-time reduction of 3SAT to the maximum
matroid-greedoid intersection problem. Therefore we have the
following.

\begin{theorem}
The maximum intersection problem for a matroid family and a greedoid
family that are given by uniform polynomial-time oracles is $NP$-hard.
\end{theorem}

Also the \emph{maximum weight} matroid-greedoid intersection problem
is $NP$-hard since maximum matroid-greedoid intersection problem is
its special case.  In this problem one should find, given integer
weights $w(x)$ for $x \in S_h$, a set $X \in F_h \cap G_h$ such that
$\sum_{x \in X} w(x)$ is maximum.

\section{Inapproximability \label{s:approx}}

As the maximum matroid-greedoid intersection problem is a maximization
problem whose exact solution turned out to be $NP$-hard, it is of
interest to see whether or not an \emph{approximation algorithm} with
a performance guarantee is possible.  An approximation algorithm would
find an element in the intersection of the matroid and the greedoid
which is not necessarily the largest one.

Following the standard approach (see e.g. \cite{ac-ca,ckst-sac}), we
say that maximization problem is \emph{polynomial-time approximable
within} $r$ where $r$ is a function from $\mathbb{N}$ to $\mathbb{Q}$
if there is a polynomial-time algorithm that finds for each instance
$x$ of the problem a feasible solution with value $c(x)$ such that
\begin{displaymath}
\frac{c_{Max} (x)}{c(x)} \leq r(|x|)
\end{displaymath}
where $c_{Max}(x)$ is the largest possible value (the optimal value)
of a feasible solution of $x$. The performance ratio of such an
approximation algorithm is bounded by the performance guarantee $r$.

\begin{theorem} \label{t:ci}
The maximum intersection problem for a matroid family and a greedoid
family with domains $\left\{ S_h : h \in H \right\}$, given by uniform
polynomial-time oracles, is not polynomial-time approximable within
$|S_h|^{1-\epsilon}$ for any fixed $\epsilon >0$, unless $P=NP$.
\end{theorem}
\begin{proof}
Assume that for some $\epsilon > 0$, the maximum matroid-greedoid
intersection problem is polynomial-time approximable within
$|S_h|^{1-\epsilon}$.  We show that then we can solve 3SAT in
polynomial time.

As in the proof of Lemma~\ref{l:bfmgi}, let $h$ again be a boolean
formula with $n$ variables in $3$-conjunctive normal form.  Now set
$S_h$ contains in addition to the truth assignment symbols
$t_1,f_1,\ldots ,t_n,f_n$ also some indicator elements ${p_i, 1 \leq i
\leq I(\epsilon )}$.  Here the number of indicators, $I(\epsilon )$,
depends on $\epsilon$ as will be shown below.  The indicators are
needed for padding the elements of the matroid and the greedoid such
that the maximum intersection becomes for a satisfiable $h$
sufficiently larger than for a non-satisfiable $h$.

We again construct a matroid $(S_h,F_h)$ and a greedoid $(S_h,G_h)$ as
follows.

The subset collection $F_h$ contains all subsets of $S_h$ that do not
contain both $t_i$ and $f_i$ for any $1\leq i \leq n$.  It is again
clear, that $(S_h,F_h)$ satisfies properties (M1) and (M2).  As
regards (M3), let $X,Y \in F_h$ such that $|X|>|Y|$.  If there is some
indicator $x$ in $X \setminus Y$, then $Y \cup \left\{ x \right\} \in
F_h$.  Otherwise $X$ must contain more truth assignment symbols than
$Y$.  Then there must be index $i$ such that either $t_i$ or $f_i$,
call it $x$, belongs to $X$ but neither of $t_i$ and $f_i$ belongs to
$Y$.  Then $Y \cup \left\{ x \right\} \in F_h$.  Thus $(S_h,F_h)$ is a
matroid.

The subset collection $G_h$ consists of three groups.  Groups A and B
are exactly same as in the construction of Lemma~\ref{l:bfmgi}.  Hence
the sets in groups A and B do not contain any indicator elements.
Group C consists of the sets of size $n$ in groups A and B, padded
with indicators in all possible ways.  That is, if $X \in A$ or $X \in
B$ such that $|X|=n$ and $Q$ is a non-empty subset of $\left\{ p_1,
\ldots , p_{I(\epsilon )} \right\}$, then $X \cup Q$ belongs to group
C.

To verify that $(S_h,G_h)$ is a greedoid, property (M1) clearly holds.
To verify (M3), let $X,Y \in G_h, |X|>|Y|$ and consider the following
cases.
\begin{enumerate}
\item
If $|Y|<n$ then there is a truth assignment symbol $x \in X \setminus
Y$ such that $Y \cup \left\{ x \right\}$ belongs to group A or to
group B as shown in the proof of Lemma~\ref{l:bfmgi}.
\item
If $|Y| \geq n$ then there is an indicator $x \in X \setminus Y$ and
thus $Y \cup \left\{ x \right\}$ belongs to group C.
\end{enumerate}

By our construction, the boolean formula $h$ is satisfiable if and
only the largest element in $F_h \cap G_h$ is of size
$|S_h|-n=I(\epsilon )+n$: The matroid contains all subsets of $S_h$
corresponding to truth assignments with all possible paddings with the
padding elements.  The greedoid contains all satisfying truth
assignments but no unsatisfying truth assignment.  Thus if $h$ is
satisfiable then the largest element in $F_h \cap G_h$ is of size
$|S_h|-n$ consisting of a satisfying truth assignment and all padding
elements.  If $h$ is not satisfiable then the greedoid does not
contain any complete truth assignment.  Since the padding elements can
occur in a set $X \in G_h$ only if $\left|X \cap \left\{
t_1,f_1,\ldots,t_n,f_n \right\}\right| \geq n$, the size of the
largest element is at most $n-1$.

Let now $I(\epsilon )=\left( 2n \right)^{1/\epsilon} -2n$.  Thus
$|S_h|= \left( 2n \right)^{1/\epsilon}$.  To test the satisfiability
of $h$ we use the approximation algorithm to find a approximately
largest element of $F_h \cap G_h$.  Let $c$ be the size of this
element.  If $h$ is not satisfiable then certainly $c<n$.  On the
other hand, if $h$ is satisfiable, then the largest element of $F_h
\cap G_h$ is of size $|S_h|-n$.  Therefore
\begin{displaymath}
\frac{|S_h|-n}{c} \leq |S_h|^{1-\epsilon}.
\end{displaymath}

But then
\begin{displaymath}
c \geq \frac{|S_h|-n}{|S_h|^{1-\epsilon}} \geq \frac{|S_h|}{2|S_h|^{1-\epsilon}}=\frac{|S_h|^\epsilon}{2} = n
\end{displaymath}
where the second inequality follows from $|S_h| \geq 2n$.  Hence $c
\geq n$ if $h$ is satisfiable and $c<n$ if it is not.  We have a
polynomial-time satisfiability test because $I(\epsilon )$ is a
polynomial in $n$ and hence in $|h|$ when $\epsilon$ is fixed, and
therefore the matroid family $\left\{ (S_h,F_h)_{h \in \mathrm{3CNF}}
\right\}$ and the greedoid family $\left\{ (S_h,G_h)_{h \in
\mathrm{3CNF}} \right\}$ can be represented by uniform oracles whose
run times are polynomial in $|S_h|$, hence in $|h|$.
\end{proof}

It is obvious that the maximum weight matroid-greedoid intersection
problem is at least as difficult as the maximum matroid-greedoid
intersection problem.  The approximability gap between these two
problems turns out to be exponential: a special case of the maximum
weight matroid-greedoid intersection, weighted greedoid maximization,
turns out to be inapproximable within $2^{|S_h|^k}$ for any fixed $k$.
(Note that instead of the bound $2^{|S_h|^k}$, any function computable
in time polynomial in $|S_h|$ would be suitable.  The explicit
function $2^{|S_h|^k}$ was chosen for the sake of concreteness.)

The \emph{weighted greedoid maximization problem} for a greedoid
family $\left\{ (S_h,G_h)_{h \in H} \right\}$ is, given an index $h$
and weights $w(x)$ for $x \in S_h$, to find a set $X \in G_h$ such
that the weight of the set $X$,
\begin{displaymath}
w(X)=\sum_{x \in X} w(x),
\end{displaymath}
is maximum.  The problem is known to be $NP$-hard \cite{kv-co}.

\begin{theorem} \label{t:wg}
The weighted greedoid maximization problem is not polynomial-time
approximable within $2^{|S_h|^k}$ for any fixed $k>0$, unless $P=NP$.
\end{theorem}
\begin{proof}
Assume that for some $k > 0$, the weighted greedoid maximization
problem is polynomial-time approximable within $2^{|S_h|^k}$.  We show
that then we can solve 3SAT in polynomial time.

Let $h$ be a boolean formula with $n$ boolean variables.  Then let
$S_h$ be the set $\left\{ t_1,f_1 , \ldots, t_n,f_n,1 \right\}$ where
$t_i$ and $f_i$ correspond to true and false truth assignments for the
$i$th boolean variable of the formula $h$, respectively, and $1$ is an
indicator element for satisfying truth assignments.  The set
collection $G_h$ consists of two groups.  The first group consists of
all subsets of $S_h \setminus \left\{ 1 \right\}$ of size at most
$n+1$.  The second group consists of the subsets of $S_h$ that contain
$1$ and represent satisfying truth assignments of $h$ and hence are of
size $n+1$.

Clearly (M1) and (M3) hold and thus $(S_h,G_h)$ is a greedoid.

We give weights to the elements of $S_h$ as follows.  The indicator
$1$ has weight $(n+1)2^{|S_h|^k}-n+1$ and the symbols $t_1,f_1, \ldots
, t_n,f_n$ have weight 1 each.  Then the maximum weight set $X \in
G_h$ has weight $n+1$ if the formula is unsatisfiable and
$(n+1)2^{|S_h|^k}+1$ otherwise.  Since
\begin{displaymath}
\frac{(n+1)2^{|S_h|^k}+1}{n+1} > 2^{|S_h|^k},
\end{displaymath}
we could separate these two cases using the approximation algorithm
and thus $P$ would be equal to $NP$.
\end{proof}

The weighted greedoid maximization problem is a special case of the
maximum weight matroid-greedoid intersection problem since we can
choose the matroid's set collection $F_h$ to be a superset of the
greedoid's set collection $G_h$, e.g., $F_h= \left\{ X : X \subseteq
S_h \right\}$.

\begin{corollary} \label{t:wmg}
The maximum weight intersection problem for a matroid family and a
greedoid family with domains $\left\{ S_h : h \in H \right\}$, given
by uniform polynomial-time oracles, is not polynomial-time
approximable within $2^{|S_h|^k}$ for any fixed $k>0$, unless $P=NP$.
\end{corollary}

Note that the maximization problem for unweighted greedoids is
trivially in $P$.

\section{Fixed-parameter intractability \label{s:fpt}}
Parameterized complexity contemplates the computational complexity of
decision problems when some parameters of the problems, e.g. the
number of vertices in a vertex cover or the maximum length allowed for
a shortest common supersequence, are fixed \cite{df-pc}.  This is
motivated by the observation that many problems have natural
parameters that are quite small in practical applications of the
problem.  A \emph{parameterized language}, representing the positive
instances of the parameterized (decision) problem, is a set $L
\subseteq \Sigma^* \times \mathbb{N}$ where $\Sigma$ is the input
alphabet and $\mathbb{N}$ is the set of parameters.

A parameterized language $L$ is said to be \emph{fixed-parameter
tractable} (FPT) if there is an algorithm $A$ that decides whether
$(e,k) \in L$ for every instance $(e,k) \in \Sigma^* \times
\mathbb{N}$ in time $f(k)|e|^c$, where $f$ is an arbitrary function
and $c$ is a constant independent from $k$.  Parameterized languages
have a hierarchy (called the $W$-hierarchy) similar to the polynomial
hierarchy
\begin{displaymath}
FPT \subseteq W[1] \subseteq W[2] \subseteq \ldots \subseteq W[P].
\end{displaymath}
It is believed that the containments are proper \cite{df-fpt}.

A parameterized language $L$ reduces to another such language $L'$ by
a \emph{standard parameterized $m$-reduction} if there are functions
$f$ and $g$ from $\mathbb{N}$ to $\mathbb{N}$, and a function $(e,k)
\mapsto e'$ from $\Sigma^* \times \mathbb{N}$ to $\Sigma^*$, such that
$(e,k) \mapsto e'$ is computable in time $g(k)|e|^c$, and $(e,k) \in
L$ if and only if $(e',f(k)) \in L'$.

In our case $L'$ will be presented as an intersection of a matroid and
a greedoid, given by uniform polynomial-time oracles. Then the
reduction step $(e,k) \mapsto e'$ is implemented by specializing the
oracle algorithms to $(e,k)$ such that they then recognize the matroid
and the greedoid for $e'$. This specialization can take time
$g(k)|e|^c$.

The \emph{parameterized weighted circuit satisfiability} is a
fundamental $W[P]$-complete problem.  This problem asks, given a
boolean circuit $h$ and a positive integer $k$, to decide whether or
not there exists a satisfying truth assignment of weight $k$, i.e., a
satisfying truth assignment with exactly $k$ variables set to $\true$.

The \emph{parameterized intersection problem} for a matroid family
$\mathcal{F}$ and a greedoid family $\mathcal{G}$ is to decide, given
an index $h \in H$ and a parameter $k$, whether or not there exists a
set $X \in F_h \cap G_h$ such that $|X|=k$.

The \emph{dual parameterized intersection problem} for a matroid
family $\mathcal{F}$ and a greedoid family $\mathcal{G}$ is to decide,
given an index $h \in H$ and a parameter $k$, whether or not there
exists a set $X \in F_h \cap G_h$ such that $|X|=|S_h|-k$.

We will show, by reduction from the parameterized weighted circuit
satisfiability, that these natural parameterizations of the maximum
matroid-greedoid intersection problem are $W[P]$-hard.  We consider
only the above versions of these problems where the solution is
required to be of certain size.  As the decision version of the
maximum matroid-greedoid intersection problem is in $NP$, the (dual)
parameterized matroid-greedoid intersection problems with inequality
constraints on the size of the solution are fixed-parameter
polynomial-time equivalent to the (dual) parameterized
matroid-greedoid intersection problem (with equality constraint on the
size of the solution) \cite[page 51]{df-pc}.

\begin{theorem}
The parameterized intersection problem for a matroid family and a
greedoid family, given by uniform polynomial-time oracles, is
$W[P]$-hard. \label{t:mgi}
\end{theorem}
\begin{proof}
We reduce the parameterized weighted circuit satisfiability to the
parameterized matroid-greedoid intersection problem as follows.

Let $\mathcal{C}$ denote the set of boolean circuits and let $h=(e,k)
\in \mathcal{C} \times \mathbb{N}$ be an instance of the parameterized
weighted circuit satisfiability problem. The circuit $e$ has $n$
variables.  To construct the corresponding matroid $(S_h,F_h)$ and
greedoid $(S_h,G_h)$, we let the set $S_h$ consist of symbols
$t_1,\ldots , t_n,1,d_1, \ldots ,d_{|e|}$ where $|e|$ is the size of
the circuit $e$ in some fixed encoding scheme.  The symbol $t_i$
denotes that the $i$th variable of $e$ is $\true$.  Symbol $1$ is an
indicator element for satisfying truth assignments of $k$ $\true$
variables.  Symbols $d_1,\ldots ,d_{|e|}$ are padding elements only
needed to make $S_h$ large enough such that the value of the circuit
$e$ can be computed in polynomial time in $|S_h|$.

The set collection $F_h$ consists of all subsets of $\left\{t_1,
\ldots ,t_n,1 \right\}$ of size at most ${k+1}$ containing a maximum
of $k$ symbols $t_i$.  Clearly the matroid properties (M1), (M2), and
(M3) hold.

The set collection $G_h$ consists of three groups.  Group A consists
of the subsets of $\left\{ t_1, \ldots ,t_n \right\}$ of size at most
$k$ representing truth assignments of maximum $k$ $\true$ variables.
Group B consists of the subsets of $\left\{ t_1, \ldots ,t_n ,1
\right\}$ of size ${k+1}$ representing truth assignments of weight $k$
that satisfy $e$.  Hence each member $X$ of the group B contains
element $1$ and elements $t_i$ for the $k$ variables with value
$\true$ in an assignment satisfying $e$.  Group C consists of the
subsets of $\left\{t_1, \ldots ,t_n \right\}$ of size ${k+1}$.

It is immediate that (M1) holds.  To verify (M3), let $X, Y \in G_h$
such that $|X|>|Y|$.
\begin{enumerate}
\item
If $|Y|<k$ then there is $x \in X \setminus Y$ such that $Y \cup
\left\{ x \right\}$ is in group A.
\item
If $|Y|=k$ then there is $x \in X \setminus Y$ such that $Y \cup
\left\{ x \right\}$ is in group B or in group C.
\end{enumerate}
Thus $(S_h,G_h)$ is a greedoid.

We now have families $\mathcal{F}= \left\{ (S_h,F_h)_{h \in
\mathcal{C} \times \mathbb{N}} \right\}$ and $\mathcal{G}= \left\{
(S_h,G_h)_{h \in \mathcal{C} \times \mathbb{N}} \right\}$ that
obviously can be given by uniform polynomial-time oracles. Given
$h=(e,k)$ the oracles check memberships in $F_h$ and $G_h$ in time
polynomial in $|e|$.  As $|S_h|= \Theta (|e|)$, because of the padding
elements, these times are polynomial in $|S_h|$, too.

The boolean circuit $e$ has a satisfying truth assignment of weight
$k$ if and only if the group B is non-empty, that is, if and only if
there is a set $X \in F_h \cap G_h$ such that $|X|=k+1$: the matroid
ensures that the solution $X$ sets at most $k$ variables $\true$ and a
set $X \in G_h$ of size $k+1$ contains $1$ only if the truth
assignment corresponding to the $k$ $t_i$'s in $X$ satisfy the circuit
$e$.  In the standard parameterized $m$-reduction we may thus choose
$f(k)=k+1$.  Moreover, the reduction $h=(e,k) \mapsto (S_h,F_h,G_h)$
is done by specialization to $h=(e,k)$ of the oracles.  This can
obviously be done in time $O(|e|+k)=O(|e|)$.  Hence we may select
$g(k)=$constant.

Thus, the parametrized matroid-greedoid intersection problem is
$W[P]$-hard.

\end{proof}

\begin{theorem}
The dual parameterized intersection problem for a matroid family and a
greedoid family, given by uniform polynomial-time oracles, is
$W[P]$-hard.
\end{theorem}
\begin{proof}
We reduce the parameterized weighted circuit satisfiability to the
dual parameterized matroid-greedoid intersection problem as follows.

Let again $h=(e,k) \in \mathcal{C} \times \mathbb{N}$ be an instance
of the parameterized weighted circuit satisfiability where $e$ has $n$
variables.

The set $S_h$ consists of $f_1,\ldots , f_n,1,d_1, \ldots ,d_{|e|}$.
The symbol $f_i$ denotes that the $i$th variable is set to be $\false$
and $1$ is an indicator element for satisfying truth assignments.  The
symbols $d_1,\ldots ,d_{|e|}$ ensure that the value of the circuit can
be computed in time polynomial in $|S_h|$. Unlike in the proof of
Theorem~\ref{t:mgi}, the padding symbols $d_1, \ldots , d_{|e|}$ are
now used in the subset collections $F_h$ and $G_h$.

The subset collection $F_h$ consists of all subsets of $S_h$
containing at most {$n-k$} symbols $f_i$.  Clearly the matroid
properties (M1), (M2), and (M3) hold.

The subset collection $G_h$ consists of four groups.  The first group
A' consists of all subsets of $\left\{ d_1 , \ldots , d_{|e|}
\right\}$.  The second group A consists of subsets $X$ of $\left\{
f_1, \ldots ,f_n , d_1, \ldots , d_{|e|} \right\}$ such that $\left\{
d_1, \ldots ,d_{|e|} \right\} \subset X$ and $|e|+1 \leq |X| \leq
|e|+n-k$, representing the truth assignments of maximum ${n-k}$
$\false$ variables.  The third group B consists of subsets $X$ of the
set $\left\{ f_1, \ldots, f_n , 1, d_1, \ldots ,d_{|e|} \right\}$ of
size $|e|+n-k+1$ representing the {$n-k$} $\false$ variables in a
truth assignment of weight $k$ that satisfies $e$.  The fourth group C
consists of subsets $X$ of $\left\{f_1,\ldots , f_n, d_1, \ldots
,d_{|e|} \right\}$ of size $|e|+n-k+1$ representing truth assignments
of $n-k+1$ $\false$ variables.  Thus $\left\{ d_1 , \ldots ,d_{|e|}
\right\} \subset X$.

It is immediate that (M1) holds for $(S_h,G_h)$.  To verify (M3), let
$X, Y \in G_h$ such that $|X|>|Y|$.
\begin{enumerate}
\item
If $|Y|<|e|$ then there is $x \in X \setminus Y$ such that $Y \cup
\left\{ x \right\}$ belongs to group A'.
\item
If $|e| \leq |Y|< |e|+n-k$ then there is $x \in X \setminus Y$ such
that $Y \cup \left\{ x \right\}$ belongs to group A.
\item
If $|Y|=|e|+n-k$ then there is $x \in X \setminus Y$ such that $Y \cup
\left\{ x \right\}$ is in group B or in group C.
\end{enumerate}
Thus $(S_h,G_h)$ is a greedoid.

The boolean circuit $e$ has a satisfying truth assignment of weight
$k$ if and only if the group B is non-empty, that is, if and only if
there is a set $X \in F_h \cap G_h$ such that $|X|=|S_h|-k$: the
matroid ensures that the solution $X$ sets at most $n-k$ variables
$\false$ and a set $X \in G_h$ of size $\left|S_h\right|-k$ contains
$1$ only if the truth assignment corresponding to the $n-k$ $f_i$'s in
$X$ satisfy the circuit $e$.  In the $m$-reduction we may hence choose
$f(k)=k$.  The rest of the proof is similar to the proof of
Theorem~\ref{t:mgi}.

Thus the dual parametrized matroid-greedoid intersection problem is
$W[P]$-hard.
\end{proof}

\section{Conclusions}
We have shown that the maximum intersection problem for a matroid
family $\left\{ (S_h,F_h)_{h \in H} \right\}$ and a greedoid family
$\left\{ (S_h,G_h)_{h \in H} \right\}$ is $NP$-hard, $W[P]$-hard and
inapproximable within $|S_h|^{1-\epsilon}$ for any fixed $\epsilon >
0$.  We have also shown that the weighted greedoid maximization is
inapproximable within $2^{|S_h|^k}$ for any fixed $k$, and thus the
weighted maximum matroid-greedoid intersection problem is
inapproximable within $2^{|S_h|^k}$ for any fixed $k$.

The maximum matroid-greedoid intersection problem is closely related
to the maximum matroid-greedoid partition problem \cite{e-mpm}.  The
\emph{maximum partition problem} for a matroid family $\left\{
(S_h,F_h)_{h \in H} \right\}$ and a greedoid family $\left\{
(S_h,G_h)_{h \in H} \right\}$ is to find, given an index $h \in H$, a
set $X= Y \cup Z, Y \cap Z = \emptyset, Y \in F_h, Z \in G_h$ such
that $|X|$ is maximum.  The $NP$-hardness and inapproximability of the
maximum matroid-greedoid intersection problem can be transformed to
show that also the maximum matroid-greedoid partition problem is
$NP$-hard and inapproximable.  The fixed-parameter (in)tractability of
the maximum matroid-greedoid partition problem is still open.

%%%%%%%%%%%%%%%%%%%%%%%%%%%%%%%%%%%%%%%%%%%%%%%%%%%%%%%%%%%%%%%%%%%%%%%%%%%%%

%
% Bibliography
%

\thispagestyle{empty}

% last page for page counting
\label{lastpage}

%
% Appendices
%

% 
% Back cover
%
%
%\cleardoublepage
%\pagestyle{empty}
%\makeemptypage
%\backcover
\end{document}